\documentclass[aps,prb,reprint,amsmath,amssymb,superscriptaddress,showpacs,floatfix]{revtex4-2}
\usepackage{graphicx}
\usepackage{dcolumn}
\usepackage{bm}
\usepackage[colorlinks, allcolors=blue]{hyperref}
\usepackage[usenames, dvipsnames]{color}
\usepackage{float}
\usepackage{lipsum}
\usepackage{titlesec}
\titlespacing\section{0pt}{12pt plus 4pt minus 4pt}{1pt plus 20pt minus 2pt}
\usepackage{xcolor}
\usepackage{tabularx}
\usepackage{amsmath}
\usepackage{comment}
\usepackage{afterpage}
\usepackage{placeins}
\usepackage{booktabs}
\usepackage{multirow}
\usepackage{array}
\usepackage{setspace}
\graphicspath{{Figs/}}
\usepackage{siunitx}
\usepackage{hhline}
\usepackage{xfrac}
\usepackage{float,graphicx}
\usepackage{mathtools}
\usepackage{listings}
\usepackage{amssymb}
\usepackage[normalem]{ulem}
\usepackage{titlesec}
\usepackage{amsfonts}
\usepackage[version=4]{mhchem}

\usepackage{epstopdf}
\usepackage{soul}
\catcode`@11
\def\seceqaa{\@addtoreset{equation}{section}
principles\def\theequation{A\arabic{equation}}}
\def\seceqbb{\@addtoreset{equation}{section}
\def\theequation{B\arabic{equation}}}
\def\seceqcc{\@addtoreset{equation}{section}
\def\theequation{C\arabic{equation}}}
\def\seceqdd{\@addtoreset{equation}{section}
\def\theequation{D\arabic{equation}}}
\def\seceqee{\@addtoreset{equation}{section}
\def\theequation{E\arabic{equation}}}
\def\seceqff{\@addtoreset{equation}{section}
\def\theequation{F\arabic{equation}}}
\def\seceqgg{\@addtoreset{equation}{section}
\def\theequation{G\arabic{equation}}}
\def\seceqhh{\@addtoreset{equation}{section}
\def\theequation{H\arabic{equation}}}
\catcode`@11
\usepackage{setspace}

\begin{document}
\makeatletter
\let\origaddcontentsline\addcontentsline
\let\addcontentsline\@gobblethree
\makeatother
PHYSICAL REVIEW B \textbf{113}, 045114 (2026)

\title{Emergence of a hidden-order phase well below the charge density wave transition in a topological Weyl semimetal (TaSe$_4$)$_2$I} 
\setcounter{footnote}{1}
\author{Sk Kalimuddin}
\affiliation{School of Physical Sciences, Indian Association for the Cultivation of Science, Jadavpur, Kolkata 700032, India}
\author{Sudipta Chatterjee}
\affiliation{Department of Condensed Matter and Materials Physics, S. N. Bose National Centre for Basic Sciences, JD Block, Sector III, Salt Lake, Kolkata 700106, India}
\author{Arnab Bera}
\affiliation{School of Physical Sciences, Indian Association for the Cultivation of Science, Jadavpur, Kolkata 700032, India}
\author{Satyabrata Bera}
\affiliation{School of Physical Sciences, Indian Association for the Cultivation of Science, Jadavpur, Kolkata 700032, India}
\author{Deep Singha Roy}
\affiliation{School of Physical Sciences, Indian Association for the Cultivation of Science, Jadavpur, Kolkata 700032, India}
\author{Soham Das}
\affiliation{School of Physical Sciences, Indian Association for the Cultivation of Science, Jadavpur, Kolkata 700032, India}
\author{Tuhin Debnath}
\affiliation{School of Physical Sciences, Indian Association for the Cultivation of Science, Jadavpur, Kolkata 700032, India}
\author{Ashis K. Nandy}
\affiliation{School of Physical Sciences, National Institute of Science Education and Research Bhubaneswar, An OCC of Homi Bhabha National Institute, Khurda Road, Jatni, Odisha 752050, India}
\author{Shishir K. Pandey}
\email{Contact author: shishir.kr.pandey@gmail.com}
\affiliation{Department of General Sciences, Birla Institute of Technology and Science, Pilani-Dubai Campus, Dubai International Academic City, Dubai 345055, UAE}
\affiliation{Department of Physics, Birla Institute of Technology and Science, Pilani, Hyderabad Campus, Jawahar Nagar, Kapra Mandal, Medchal District, Telangana 500078, India}

\author{Mintu Mondal}
\email{Contact author: sspmm4@iacs.res.in}
\affiliation{School of Physical Sciences, Indian Association for the Cultivation of Science, Jadavpur, Kolkata 700032, India}

\date{\today}% It is always \today, today,

\begin{abstract}
The emergence of a charge density wave (CDW) in a Weyl semimetal-- a correlated topological phase, is exceptionally rare in condensed matter systems. In this context, the quasi-one-dimensional type-III Weyl semimetal (TaSe$_4$)$_2$I undergoes a CDW transition at $T_{\mathrm{CDW}} \approx 263$~K, providing an exceptional platform to investigate correlated topological CDW states. Here, we uncover an additional hidden-order phase transition at $T^* \sim 100$ K, well below the CDW onset, using low-frequency resistance noise spectroscopy, electrical transport, and thermoelectric measurements. This transition is characterized by a sharp enhancement in the noise exponent ($\alpha$) and variance of resistance fluctuations. Analysis of higher-order statistics of resistance fluctuations reveals the correlated dynamics underlying the transition. A pronounced anomaly in the Seebeck coefficient near $T^*$ further suggests a Fermi surface reconstruction. First-principles calculations reveal a structural distortion from the high-symmetry $I422$ phase to a low-symmetry $C2$ phase, via an intermediate $I4$ symmetry. This leads to renormalization of the electronic structure near the Fermi level and opening of a bandgap in the hidden-order phase. These findings demonstrate a previously unidentified correlated phase transition in the topological CDW-Weyl semimetal (TaSe$_4$)$_2$I, enriching the phase diagram of this material and establishing it as an ideal platform for studying intertwined electronic and structural orders.
\end{abstract}

\maketitle

\section{Introduction}
The emergence and coexistence of distinct ordered states constitute a hallmark of strongly correlated systems, and deciphering their microscopic mechanisms lies at the forefront of condensed matter research. Symmetry breaking in such systems often drives the formation of new ordered states, manifested by the opening of an energy gap near the Fermi level ($E_F$) \cite{gruner2018density,tinkham1996superconductivity,lee1973conductivity}. This gap is typically proportional to the magnitude of the corresponding order parameter. One well-known example is the charge density wave (CDW) phases in low-dimensional materials, characterized by periodic modulation of conduction electrons accompanied by a corresponding lattice distortion \cite{peierls1955quantum,gruner2018density,monceau2012electronic}. CDW states often coexist with other emergent quantum phases, including superconductivity \cite{Superconducting_CDW,Superconducting_CDW_TaS2}, and ferromagnetism \cite{FerromagneticCDW}. Furthermore, CDW systems serve as a versatile ground for realizing intriguing correlated phenomena such as the quantum anomalous Hall effect \cite{AHall_CDW,FQAHall_CDW_PRB}, axionic electrodynamics \cite{AxionGooth2019}, valley-selective transport \cite{Valley_Polarization}, and critical slowing down near phase transitions \cite{Dynamical_Slowing_DownPRL}.

The fundamental collective excitations of the CDW state, amplitudon and phason, are manifestations of spontaneous symmetry breaking in the ordered phase \cite{gruner2018density,Gedik_Collective}. Recently, the dynamics of these collective modes have been extensively investigated using a variety of experimental techniques \cite{Dynamical_Slowing_DownPRL,kim2023observation,Fahad.CollectiveMode.PRL}. Owing to the finite gap in its dispersion, the amplitude mode contributes negligibly at low temperatures, except in the vicinity of the transition temperatures. By contrast, the phason is expected to be a gapless Goldstone mode; however, in real materials, impurities and defects lift this degeneracy, opening a finite gap and endowing the phason with mass \cite{kim2023observation}. The resulting low-energy excitations, particularly those associated with mass-acquired phasons and hidden CDW orders, are of fundamental interest \cite{Fahad.CollectiveMode.PRL,demsar1999single,Gedik_Collective}. 

The broad family of transition metal tetrachalcogenides that have the generic formula (MX$_4$)$_n$Y (where M = Nb, Ta; X = S, Se; Y = halogens such as Br, I; and n = 2, 3, 10/3) serves as quasi-one-dimensional model systems (quasi-1D) to study the role of filling the metal band $d_{z^2}$ \cite{gressier1984characterization,Monceau1982,Roucau_1984,PhysRevB.94.104113,MMondalnTSI.PRB,Balandin2024,BALANDIN202274Review} giving rise to unconventional ground states, such as 1D band insulator \cite{1DbandinsulatorPhysRevLett.84.1272}, type III Weyl phase \cite{li2021typeIIIWeyl}, which hosts giant dielectric response \cite{GiantdielectricPhysRevLett.96.046402}, CDW state with boundary modes \cite{Zahid_Nat.Phy} and so on. The highly correlated response of the ordered collective CDW modes in these materials can also give rise to complex low-energy dynamics.

Among them, (TaSe$_4$)$_2$I (TSI2 in short) is the only type III CDW Weyl semimetal having a body-centered tetragonal crystal structure (space group-$I422$) at room temperature and undergoes the CDW transition near $T_{CDW} \sim 263~K$. As a rare Weyl-CDW material, TSI2 has recently sparked interest due to its potential to realize a so-called dynamical axion insulator state \cite{AxionGooth2019}, a correlated topological phase in which the CDW phase mode becomes an analog of the predicted axial field in high-energy physics. However, this interpretation is currently the subject of active debate \cite{MonceauAPL2022,Zahid_Nat.Phy}. On the other hand, a recent study of ultrafast X-ray scattering suggested that the CDW amplitude mode is more than merely electronic; it also involves lattice degrees of freedom\cite{Fahad.CollectiveMode.PRL}. Furthermore, TSI2 has also been observed to be an exceptional CDW candidate for the achievement of a stable 1D atomic chain limit, amporphorization superconductivity, and photodetection \cite{wei2024polarization,li2024ultrabroadband,an2020long,Balandin2024}. These investigations unequivocally show that the topology and the electronic and lattice degrees of freedom of the collective CDW modes are strongly correlated in this compound. Despite this, all previous investigations in TSI2 have focused mainly on the $T_{CDW} \sim 263~K$ regime, while the low-temperature investigations of these correlated CDW collective modes are still very scarce.

Despite the wealth of extensive research on the CDW transition near $T_{CDW}$, the low-temperature regime of TSI2 remains largely unexplored, particularly concerning the nature of its correlated excitations and potential emergent orders. This unexplored region is especially intriguing, as subtle symmetry breakings or secondary instabilities—often hidden at high temperatures—can manifest more clearly at low temperatures. To probe such elusive low-energy dynamics and identify possible hidden phases, it is essential to employ sensitive techniques beyond conventional transport measurements.

In this regard, low-frequency resistance noise spectroscopy has been demonstrated to be a sensitive tool for identifying phase transitions \cite{chadniPhysRevLett.102.025701,majhi2021diffused,Kalimuddin.PRL,KaziAminPRL,Ghosh2023}. Near the transition, one can typically expect a divergent growth of the noise exponent ($\alpha$), resistance fluctuations $\left(\frac{\langle\delta R^2\rangle}{\langle R^2\rangle}\right)$, and relaxation time ($\tau$), a phenomenon often referred to as critical slowing down \cite{Dynamical_Slowing_DownPRL,Satyaki_CriticalPRL}. The appearance of critical slowing down near the phase transition is a much more general phenomenon, as observed in first-order transitions \cite{Satyaki_CriticalPRL}, second-order transitions \cite{Kalimuddin.PRL}, and also in dynamical systems \cite{Dynamical_Slowing_DownPRL}. Therefore, its widespread occurrence makes it a hallmark of phase transitions in a wide range of correlated solid-state systems.

Our experimental results suggest that there is an additional phase transition in this compound around $T$* $\sim$ 100 K, much lower than $T_{CDW}$. We observe a divergent increase in the noise exponent ($\alpha$) and resistance fluctuations close to $T$*. The correlated nature of resistance fluctuations in the vicinity of $T$* is established by the second spectrum of resistance noise. Furthermore, temperature-dependent resistivity and Seebeck coefficient exhibit a substantial anomaly at $T$*. Our DFT calculations show that a structural transition from the high-symmetry $I422$ phase to a low-symmetry $C2$ phase induces a small bandgap ($\sim$ 0.1 eV) and significant band renormalization near the Fermi level. Group-theoretical analysis further reveals that this transition proceeds via an intermediate $I4$ phase, driven by lattice distortions, indicating a structural origin for the hidden-order phase near $T^*$. Overall, our findings suggest a hidden order phase in the type III CDW Weyl semimetal (TaSe$_4$)$_2$I close to $T$ *, well below $T_{CDW}$. 

\section{Results}% and Discussion}

\subsection{Crystal structure and electrical transport}

\begin{figure}%[h]%[t]%[h]
\centering
\includegraphics[width=1.0\columnwidth]{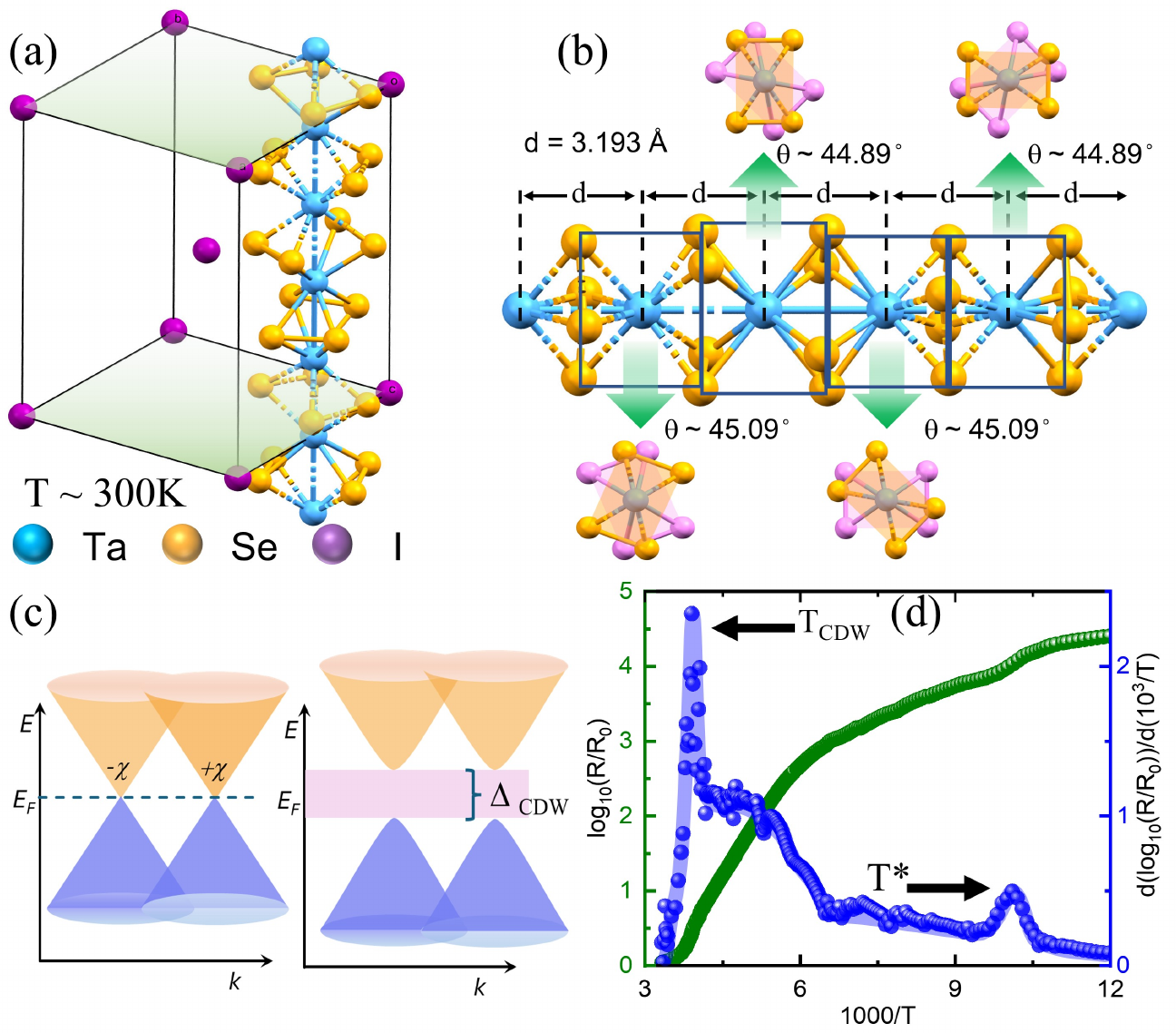} 
\caption{(a) Schematic representation of the crystal structure from single-crystal x-ray Laue diffraction at room temperature. (b) Schematic of a single (TaSe$_4$) chain along the c-axis highlights the Ta-Ta bond length and the dihedral angle among two consecutive (Se$_4$) planes. (c) A schematic of the phase transition from a Weyl semimetal into a charge density wave insulator (with Weyl cones of opposite chirality $+\chi$ and $-\chi$). (d) Temperature-dependent normalized log$_{10}(R/R_0)$ (olive green) measured in four-probe geometry [R$_0$ = R(300K)], and first derivative [d\{log$_{10}$(R/R$_0$)\}/d($10^3$/T)] (blue in right) vs ($10^3/T$).}
\label{Fig:Characterization}
\end{figure}

TSI2 single crystals have been grown using the chemical vapor transport (CVT) method using Iodine ($I_2$) as transport agent \cite{gressier1984characterization,AxionGooth2019}. The details of the growth process are also available in our earlier report \cite{Kalimuddin.PRL}. The crystal structure as inferred from single crystal Laue diffraction at room temperature reveals that TSI2 crystallizes in a body-centered tetragonal structure (space group \textit{I422}, no. 97) (see S1 of Supplemental Material (SM) \cite{supply}). The schematic representation of the projected tetragonal crystal structure has been shown in Fig. \ref{Fig:Characterization}(a). The unit cell of TSI2 consists of parallel chains of TaSe$_4$ units where Se$_4$ of TaSe$_4$ forms two dimers aligned in a plane, forming a rectangle. Each Ta atom is sandwiched at the center between two adjacent Se$_4$ planes, which is presented in Fig. \ref{Fig:Characterization}(b). Consecutive rectangular planes of Se$_4$ are skew rotated by $\sim45^{\circ}$. Infinitely long, well-separated, parallel skew-rotated TaSe$_4$ chains provide a quasi-1D character and substantial anisotropy in the electronic structure. 

TSI2 is currently the only known material system in which a CDW transition emerges from an underlying Weyl semimetal phase, potentially leading to the realization of an axion insulator state. In this system, the CDW—whose collective sliding mode is associated with an axion—is proposed to couple Weyl nodes of opposite chirality located at different points in the momentum space, thus breaking chiral symmetry and inducing an insulating gap, as illustrated in Fig. \ref{Fig:Characterization}(c). This partial opening of the gap in TSI2 is caused by imperfect Fermi surface nesting \cite{Shi2021}, which may give rise to a secondary order significantly lower than $T_{CDW}$. The temperature-dependent normalized log resistivity [log$_{10}(R/R_0)$] is plotted in Fig. \ref{Fig:Characterization}(d)(left). The CDW transition temperature ($T_{CDW}$) can be identified from the maxima of $\frac{d log_{10}(R)}{d(10^3/T)}$ vs ($1/T$) plot in Fig. \ref{Fig:Characterization}(d)(right). Notably, an additional, albeit weaker, resistivity anomaly is observed near $T^*$ $\sim$ 100 K, well below $T_{CDW}$, suggesting the possibility of a secondary phase transition. Although conventional transport measurements such as resistivity reveal only subtle signatures near $T^*$, low-frequency $1/f$ noise spectroscopy offers enhanced sensitivity to dynamical fluctuations, providing a more discerning probe to identify and characterize this secondary phase transition. Therefore, in the following subsection, we will primarily focus on the resistance noise measurement around $T^*$.

\subsection{Low-frequency resistance noise measurement}
The low-frequency resistance fluctuations for a constant drive current of 1$\mu$A are measured while stabilizing the temperature fluctuations to the best possible state of $\pm$2 milli-Kelvins to rule out and minimize the contribution of temperature fluctuations to resistance noise. The measurement at each temperature involves the acquisition of time traces of resistance fluctuations at a sampling frequency of 1024 Hz for 4800 seconds.

\begin{figure}%[h]
\centering
\includegraphics[width=1.0\columnwidth]{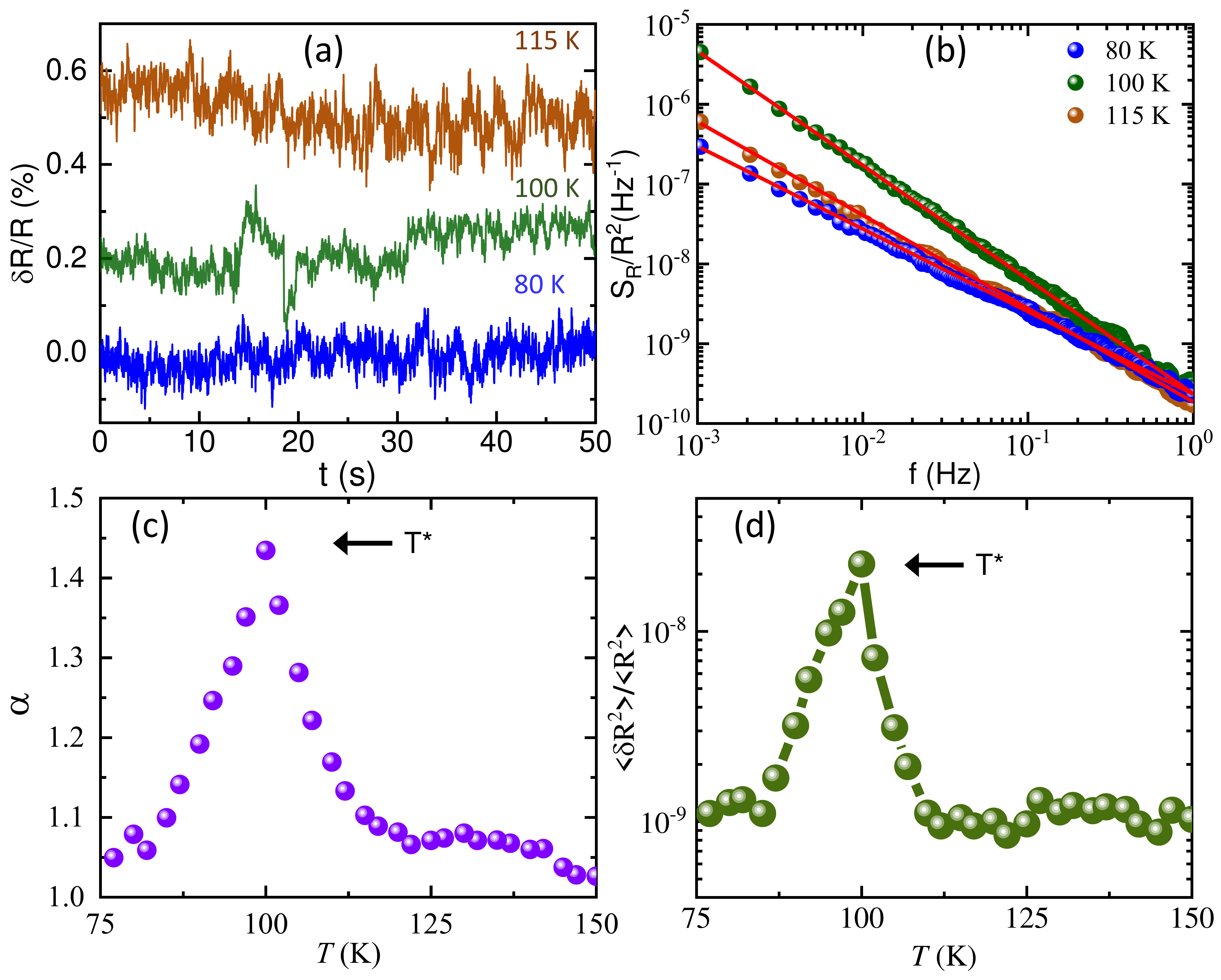} 
\caption{(a) Time series of resistance fluctuations at three distinct temperatures (normalized to absolute resistance value). (b) Plots of the normalized power spectral density of resistance fluctuations $\left(\frac{S_R(f)}{R^2}\right)$ of respective temperatures. (c) Temperature dependence of noise exponent ($\alpha$), where S$_R$(f) $\sim f^{-\alpha}$. (d) The relative noise variance of resistance fluctuations as a function of temperature.}
\label{Fig:Noise1}	
\end{figure}
Figure \ref{Fig:Noise1}(a) shows the time traces of the resistance fluctuations for a few representative temperatures. Each data point for fluctuations has been vertically shifted by 0.2 \% for clarity. Fluctuations increase in magnitude near the $T^*$ $\sim$100~K. The power spectral density (PSD) of the digitally filtered fluctuations within a bandwidth of 1 mHz–1 Hz is numerically calculated using Equation \ref{Eq:SR} employing Welch's Periodogram method \cite{ghosh2004set,Kalimuddin.PRL}.

\begin{figure*}%[b]
\centering
\includegraphics[width=1.0\textwidth]{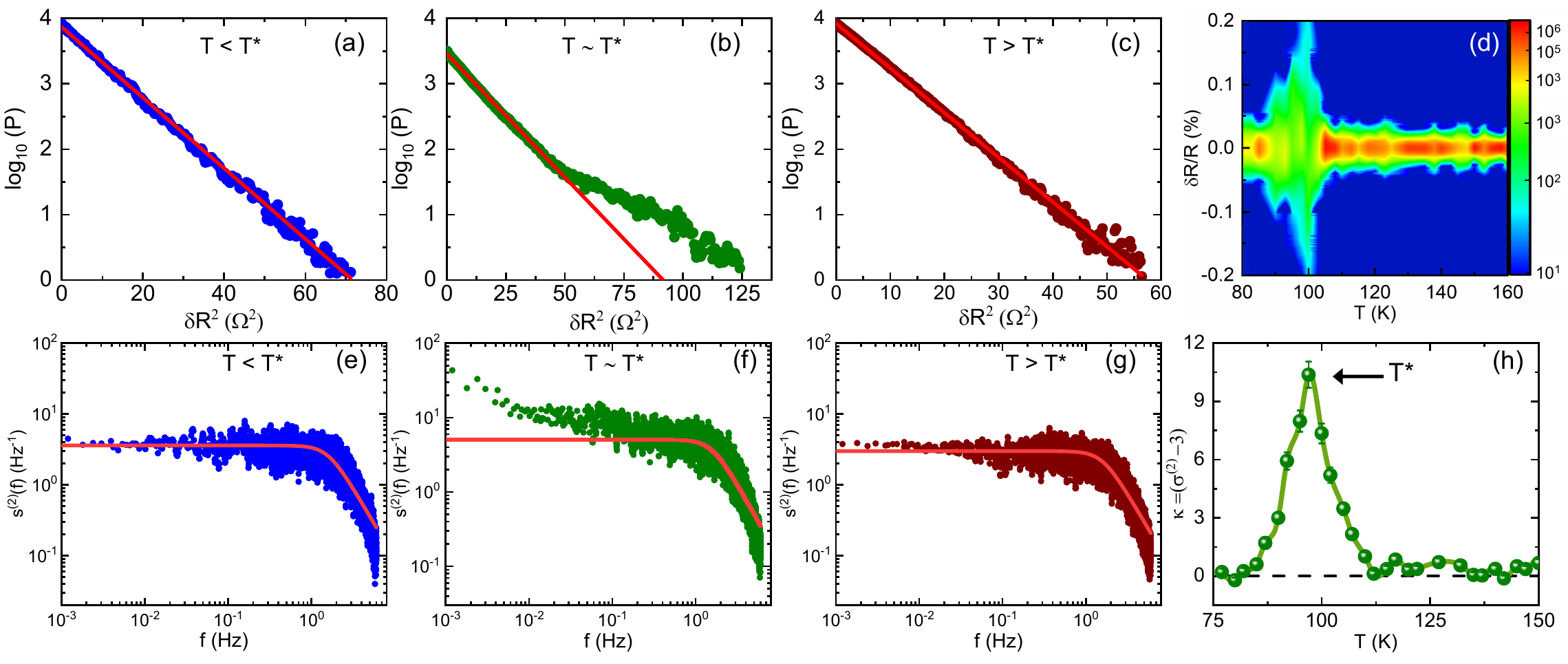}
\caption{(a)-(c) log$_{10}\left(P(\delta R)\right)$ vs $\delta R^{2}$ for temperatures $T<T^*$, $T\sim T^*$ and $T>T^*$ respectively. The solid (red) line is a guide to the Gaussian limit of the probability distribution of resistance fluctuations. The stochastic distributions of the fluctuations show extreme deviations from the normal distributions in the close vicinity of $T^*$. (d) A contour plot of PDF vs $\delta R$ with temperature. (e)-(g) Plot of normalized second spectra $s^{(2)}(f_2)$ vs frequency for temperatures $T<T^*$, $T\sim T^*$, and $T>T^*$ respectively. The solid (red) line represents the calculated Gaussian background of the second spectrum. (h) Plot of excess kurtosis ($\kappa=\sigma^{(2)}$-3) vs. temperature.}
\label{Fig:Noise2}	
\end{figure*}
\begin{equation}
S_R(f)=\lim_{T \to \infty}\Big (\frac{1}{T}\Big )\Bigg |\int_{-T/2}^{T/2} R(t)e^{-i2\pi ft}dt\Bigg |^2
\label{Eq:SR}
\end{equation}

The normalized PSD of the representative time series resistance fluctuations is plotted in Fig. \ref{Fig:Noise1}(b). The noise magnitude and the spectral character show typical $1/f^{\alpha}$ behavior \cite{CuerrentOpinion_Raychaudhuri,chadniPhysRevLett.102.025701,NbNPhysRevLett.111.197001}, where $\alpha$ is the noise exponent. The temperature-dependent plot of $\alpha$ in Fig. \ref{Fig:Noise1}(c) shows an increase upon approaching $T^*$ from $\alpha\sim1.05$ at $T$ = 85 K to $\alpha\sim1.45$ near 100~K. This rapid growth in $\alpha$ near $T^*$ suggests a strong shift of spectral weight to lower frequencies with a dominance of low-frequency/large-timescale of fluctuations \cite{Satyaki_CriticalPRL,Kalimuddin.PRL}. The relative noise variance (integrated PSD) of the fluctuations $\left( \frac{\langle\Delta{R^{2}} \rangle}{\langle R^{2}\rangle}\right)$ within the band, f$_{min}$ = 1~mHz and f$_{max}$ = 1~Hz, is estimated using the relation \ref{Eq:SR2}\cite{CuerrentOpinion_Raychaudhuri,sudiptaPhysRevB.104.155101},
\begin{equation}
\frac{\langle\Delta{R^{2}} \rangle}{\langle R^{2}\rangle} =\int_{f_{min}}^{f_{max}} \frac{S_{R}(f)}{R^{2}}df
\label{Eq:SR2}
\end{equation}

The temperature variation of $\frac{\langle\Delta{R^{2}}\rangle}{\langle R^{2}\rangle}$ in Fig. \ref{Fig:Noise1}(d) shows a rapid growth in the vicinity of $T^*$ by an order of magnitude compared to far temperature values. We have measured the low-frequency 1/f noise in another single crystal (sample II). Indeed, the signatures of the aforementioned hidden-order transition are reasonably reproduced in the second sample (see Fig. S2 of the SM\cite{supply}). Remarkably, we noticed a similar behavior in the temperature-dependence of $S_R/R^2$, $\alpha$, and $\frac{\langle\Delta{R^{2}}\rangle}{\langle R^{2}\rangle}$.

The significant jump both in $\alpha$ and $ \frac{\langle\Delta{R^{2}} \rangle}{\langle R^{2}\rangle}$ clearly indicate towards a phase transition near the $T^*$ \cite{RevModPhysWeissman,Satyaki_CriticalPRL,Kalimuddin.PRL}. It is also noteworthy that here, the $T^*$ is significantly low compared to $T_{CDW}$, suggesting the order parameter is nearly fully gapped with a long-range ordered state. Thus, the random distribution of fluctuators is not entirely valid. This indicates a possibility of correlation in the fluctuators.

In order to identify the nature of the resistance fluctuations near the $T^*$, we have calculated the probability density function (PDF) of the detrended fluctuations \cite{NbNPhysRevLett.111.197001}. For a Gaussian fluctuation with zero mean, the PDF (P($\delta R$)) is given by Eq. \ref{Eq:PDF},
\begin{equation}
 P(\delta R)=\frac{1}{\sqrt{2\pi \sigma}}e^{-\delta R^2/2\sigma^2}
\label{Eq:PDF}
\end{equation} 
Here, $\sigma^2$ represents the variance. Consequently, log$_{10}$(PDF) vs. $\delta R^2$ is a straight line, while the experimental time traces with correlation break such a pathway. Away from T$^*$, the distributions in $\delta R$ are mostly linear [Fig. \ref{Fig:Noise2}(a,c)], suggesting Gaussian characteristics in $\delta R$. However, close to $T^*$, fluctuations have strong deviations from linear distribution, indicating a non-Gaussian distribution of resistance fluctuations, as shown in Fig. \ref{Fig:Noise2}(b). In addition, the PDF broadens (growth in $\sigma$), which is a direct consequence of non-Gaussian fluctuations in the resistance noise \cite{Swastik_PhysRevLett.91.216603}as shown in \ref{Fig:Noise2}(d).

The frequency-domain determination of Kurtosis/normalized “second variance” ($k/\sigma^{(2)}$) allows an accurate determination of non-Gaussianity. We, therefore, calculate the second spectrum $\left(\Sigma^{(2)}(f_2)\right)$ from the higher-order statistics of the resistance fluctuations. $\Sigma^{(2)}(f_2)$ is a Fourier transform of the four-point correlation function, defined as $\Sigma^{(2)}(f_2)$ = $\int_{0}^{\infty} \langle \delta R^{2}(t) \delta R^{2}(t+\varsigma)\rangle \cos (2\pi f\varsigma) d\varsigma$ \cite{seidler1996non,Swastik_PhysRevLett.91.216603,chadniPhysRevLett.102.025701}. The non-Gaussian component in fluctuation is identified from the non-white character in $\Sigma^{(2)}(f_2)$. It is a measure of "spectral wandering" or fluctuations of the power spectrum itself \cite{NbNPhysRevLett.111.197001}. The normalized second spectrum is given by Eq. \ref{Eq:Sigma2}, 
\begin{equation}
 s^{(2)}(f) = \frac{\Sigma^{(2)}(f)}{\big|\int_{f_L}^{f_H} S_{R}(f)df\big|^2}
 \label{Eq:Sigma2}
\end{equation}

Figure \ref{Fig:Noise2}(e-g) shows the plot of normalized second spectra at three temperatures encompassing $T^*$. $S^{(2)}(f)$ is white (frequency independent) for temperatures away from; $T^*$ in contrast it becomes frequency dependent near $T^*$. We quantitatively estimate the excess kurtosis ($k=\sigma^{(2)}-3$) where $\left(\sigma^{(2)} = \int_{0}^{f_H - f_L} s^{(2)}(f)df\right)$ within the band [$f_L$ = $10^{-3}$~Hz and $f_H$ = 1~Hz]. 

Figure \ref{Fig:Noise2}(h) shows the temperature variation of $k$. Interestingly, $k$ exhibits a pronounced peak near the $T^*$. This unambiguously denotes correlated fluctuations close to the $T^*$, where the fluctuations are slow. However, away from the transition region, $(\sigma^{(2)}-3)$ approaches 0. This is due to the uncorrelated Gaussian nature of the fluctuations \cite{seidler1996non}. This is a direct signature of a phase transition accompanied by spatial correlation near $T^*$. Similar features of $\sigma^{(2)}$ are also evidenced near glassy freezing of electrons across the metal-insulator transition \cite{Swastik_PhysRevLett.91.216603}, Berezinskii-Kosterlitz-Thouless Transition \cite{NbNPhysRevLett.111.197001}, athermal phase transition \cite{chadniPhysRevLett.102.025701} and Peierls transition \cite{Kalimuddin.PRL}.

\subsection{Thermopower measurement}

To gain further insight into the emergence of the hidden order phase around $T^*$, we measured Seebeck (S) as a function of temperature, since it carries information about the scattering associated with soft phonons \cite{kuo2005anomalousseebeck} and finite changes in the density of states (DOS) near the Fermi surface \cite{GeometryNatMat,Sruthi2022}. The temperature-dependence of $S (T)$ is described by well known Mott's formula\cite{Mott_PRB_1969}, $S=-\frac{\pi^2 k_B^2 T}{3e} \left. \frac{d \ln \sigma(E)}{dE} \right|_{E = E_F}$, where $\sigma(E)$ represents energy-dependent electrical conductivity. Here, $k_B$, $e$, and $E_F$ represent the Boltzmann constant, electronic charge, and the Fermi energy, respectively.

\begin{figure}%[h]
\centering
\includegraphics[width=1.0\columnwidth]{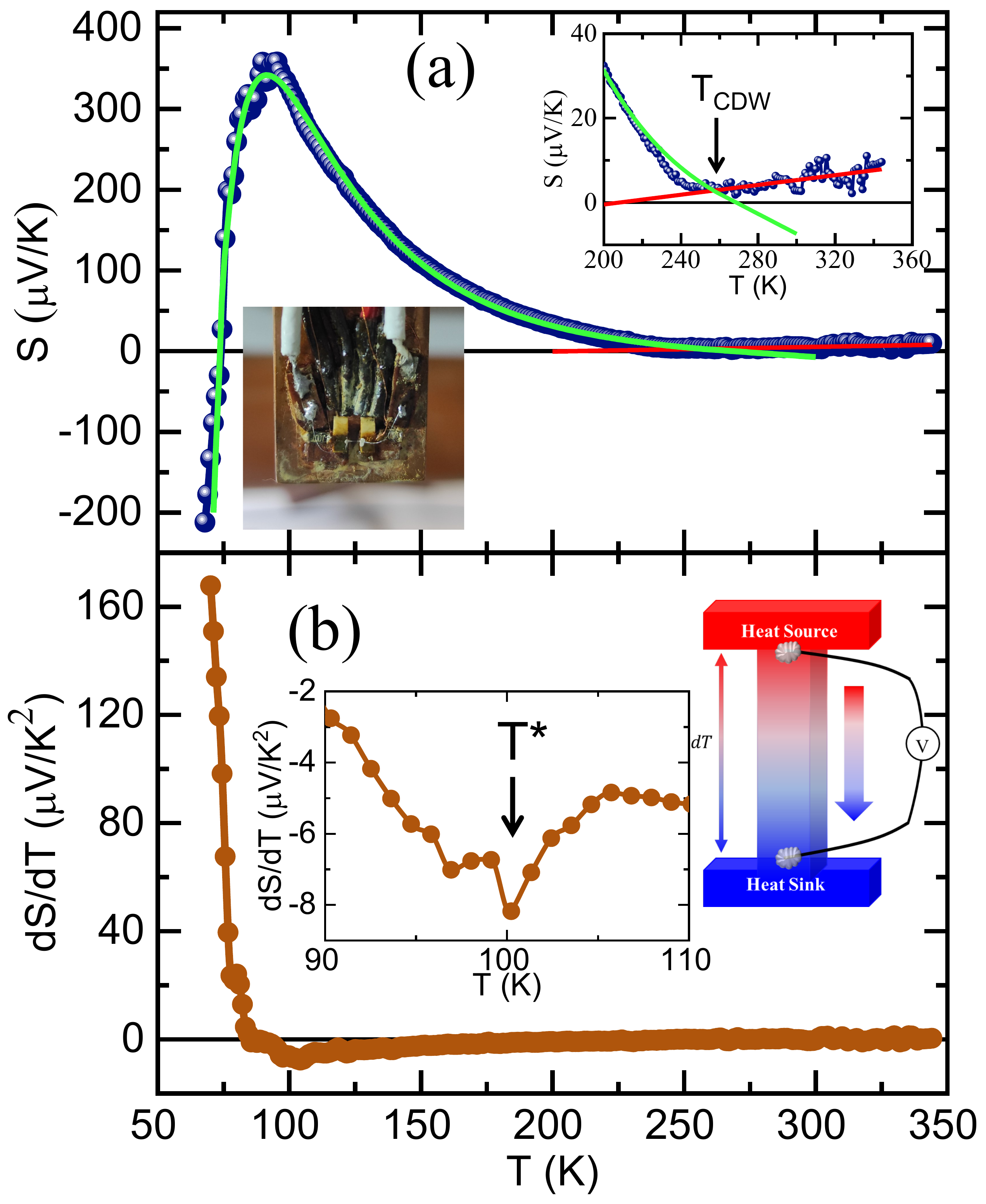} 
\caption{(a) Seebeck $S(T)$ vs. temperature along the direction of Ta chains. The solid red (semiconductor model), pink (phonon drag model), and green (phonon drag combined with metallic model using heaviside theta function). The inset shows the suspended (TaSe$_4$)$_2$I single crystal between two local heat baths. (b) Plot of the first derivative of $S(T)$ vs. temperature. The insets show the schematic representation of the steady-state gradient mode method for Seebeck measurement and an enlarged view of dS/dT close to $T^*$.}
\label{Fig:Seebeck}	
\end{figure}

The temperature-dependent Seebeck in (TaSe$_4$)$_2$I was limited down to 115~K only \cite{Bilusic2000}, which effectively captured the CDW transition and anisotropy well. Here, we extend the Seebeck coefficient measurement down to $\sim$80~K, as shown in Fig. \ref{Fig:Seebeck}(a) (for measurement methodology, see S3 of the SM\cite{supply}). The first derivative of $S(T)$ is also depicted in Fig. \ref{Fig:Seebeck}(b). S exhibits a linear and metallic behavior above the CDW transition and a nearly inverse $T$ variation below $T_{CDW}$. The crossover from linear to inverse $T$ dependence is marked by $T_{CDW}$, as shown in the inset of Fig. \ref{Fig:Seebeck}(a). At low temperatures, much below $T_{CDW}$, (TaSe$_4$)$_2$I exhibits an anomalous drop in S, suggesting a new phase transition around $T^*\sim$100~K. A similar anomaly is also observed in the temperature variation of $S(T)$ measured in another single crystal (sample II) (see S4 of the SM\cite{supply}). Here, $T^*$ is evidenced from $dS/dT$, as illustrated in the inset of Fig. \ref{Fig:Seebeck}(b). The temperature-dependence of the Seebeck coefficient in a complex, correlated system like TSI is expected to have contributions from both the diffusion of charge carriers and phonon drag\cite{MoO3_Thermopower}. The temperature-dependent Seebeck coefficient $S(T)$ in TSI2 is modeled to account for these two primary mechanisms. This is particularly important below the CDW transition at $T_{CDW}$, where strong electron-phonon coupling makes phonon drag a significant contributor.

We therefore fit $S(T)$ over the entire range to identify the contributions of different processes. Above $T_{CDW}$, TSI2 is a Weyl semimetal; consequently, $S(T)$ is linear in $T$ and can be described using the uncorrelated degenerate Fermi gas model, $\left( S(T) = -\frac{\pi^2k_B^2}{eE_F}T\right)$\cite{Mott_PRB_1969,ziman2001electrons}. Below $T_{CDW}$, the Weyl nodes become gapped due to finite lattice distortion along the TaSe$_4$ chains\cite{van2001structure}. Considering the narrow band gap of (TaSe$_4$)$_2$I below and band-crossing above T$_{CDW}$, we take into account the combined conduction for both electrons and holes (i.e, $\sigma=\sigma_c$+$\sigma_v$) and fit $S(T)= \frac{\Delta_{eff}}{T} +const$, where $\Delta_{eff}$ represents the gap \cite{fritzsche1971general}. The value of the estimated gap from Seebeck measurement is $\Delta_{eff}\sim 84$~meV.

The fit to Seebeck data deviates largely when $T$ is very close to both $T_{CDW}$ and $T^*$. The CDWs are electron-phonon coupling-driven phenomena, therefore expected to have a large contribution of strong phonon drag in $S(T)$. The complete functional form of $S(T)$ is given by \cite{bailyn1960transport,cohn1991thermoelectric}, 

\begin{equation}
S(T) =
\begin{cases} 
c+ mT, & T > T_{CDW} \\
c+ \frac{aT^2}{b+dT^3} +\frac{fT^3}{g+hT^4}, & T < T_{CDW}
\end{cases}
\label{Eq:SuppST}
\end{equation}

Here, $S_{pd,\sigma}=\frac{T^2}{b+dT^3}$ and $S_{pd,\pi}=\frac{T^3}{g+hT^4}$ are the contribution of phonon drag from $\sigma$ and $\pi$-bonds, respectively. Here, a and f are the weight factors of the phonon-drag contributions from $\pi$ and $\sigma$ orbitals of $Ta$ and $Se$, respectively. These fitting parameters (a, b, c, d, f, g, h) collectively determine the strength of bipolar conduction and the temperature profile of distinct phonon drag contributions. The obtained values of the fitting parameters, a~=~-42866.5; b = 144.9; c = -11.2; d = 0.007; f = 45227.1; g = 11058.6; h = 0.007. The cumulative contributions of the diffusive and drag contributions yield a good fit throughout the temperature range.

\subsection{First-principles calculations}

\begin{figure}
\centering
\includegraphics[width=1.0\columnwidth]{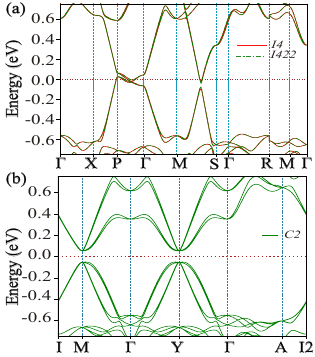}
\caption{$Ab$ $initio$ band structure for the, (a) room temperature $I422$ and mid-temperature $I4$, and (b) low temperature ($C2$) CDW phase. During the structural transition $I422$ $\rightarrow$ $I4$ $\rightarrow$ $C2$, the renormalization of the electronic structure leads to a small band gap in the low-temperature $C2$ phase as shown in (b). The contribution at the Fermi level (set to zero) is from Ta-$d$ orbitals.} 
\label{fig:bands}
\end{figure}

To further uncover the origin of the hidden phase transition, we performed first-principles band structure calculations for the three structures of TSI2 to analyze the change near the Fermi level. The plots are summarized in Fig.~\ref{fig:bands}. The band structures for the room-temperature phase $I422$ and mid-temperature (i.e. $T^*<T<T_{CDW}$) structures $I4$ look identical, as shown in Fig.~\ref{fig:bands} (a). As one can see, DFT calculations predict a semi-metallic nature, and we could not find any noticeable change in the band structure for these two structures. At the Fermi level, the contribution comes mostly from Ta-$d$ orbitals (not shown). TSI2 consists of indefinite TaSe$_4$ chains with each Ta ion being sandwiched between two consecutive $Se_4^{4-}$ rectangles. The equivalent number of $d$ electrons is $\frac{1}{2}$ in each Ta atom, and the corresponding band filling of Ta-$d$ orbitals here is $\frac{1}{4}$\cite{gressier1984electronic}. This partial filling of bands facilitates Peierl's distortion. 

Nevertheless, the CDW phase in (TaSe$_4$)$_2$I emerges below $T_{CDW}$ ($\sim$263K), the subsequent noticeable tetramerization involving Ta atoms along the chain-axis could be captured much below $T_{CDW}$\cite{van2001structure}. 

To observe the changes in the electronic structure of the $C2$ phase, we explicitly incorporate the displacement of Ta atoms along the Ta chain(along the c-axis) to depict the crystal structure change due to the CDW transition. The uniform 3.241 \AA{} Ta-Ta bonds of the original non-CDW structure change into intra-timer 3.203 \AA{} and inter-trimer and 3.279 \AA{} bonds in the CDW structure. Indeed, the CDW primitive cell (22 atoms) is $\sim$71~meV lower in energy than the original non-CDW phase. The band structure calculations for the thusly obtained CDW structure are shown in Fig.~\ref{fig:bands}(b). One can clearly see the renormalization of the band structure at the Fermi level, resulting in a small indirect band gap of $\sim$~0.1~eV at the Fermi level. The single particle activated energy gap ($E_g$) estimated from the resistivity is 216~meV, which is close to theoretical estimates of 0.1 eV. 

\begin{figure}%[b]
\begin{center}
\includegraphics[width=1.0\columnwidth]{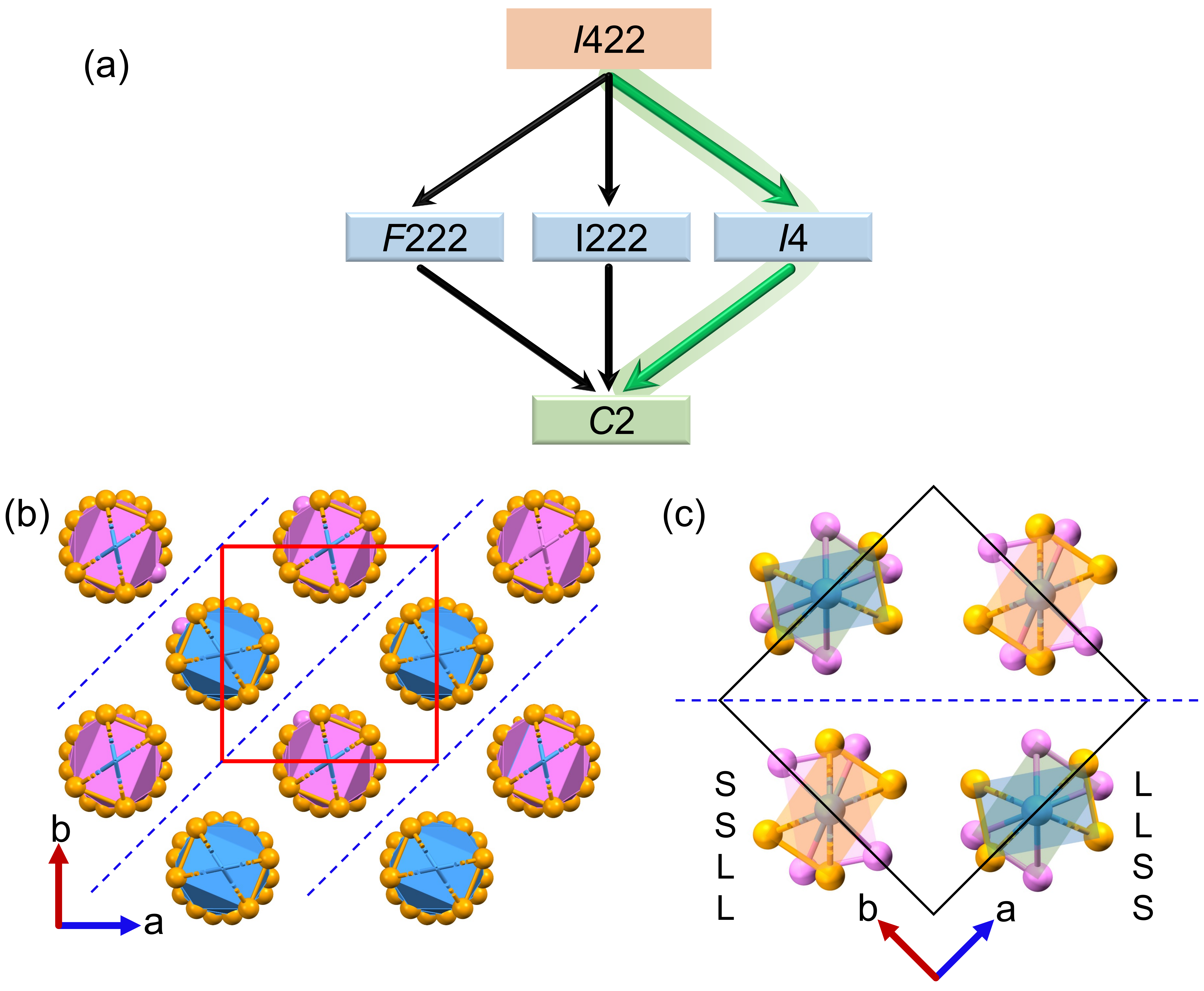}
\caption{(a) Evolution of crystal structure analyzed using the CELLSUB module~\cite{bb2} of Bilbao crystallographic server~\cite{bb1}. The green arrow indicates the energetically feasible pathway among the three possible ways for TSI2.(b-c) Schematic representation of TaSe$_4$ chains with different tetramerized structures of periodic LLSS(pink octahedra) and SSLL(blue octahedra) units, respectively. The (1,1,0) plane is marked by a dashed line. Two consecutive (1,1,0) planes sandwich TaSe$_4$ chains with an alternate sequence of LLSS and SSLL bonding sequence.} 
\label{fig:path}
\end{center}
\end{figure}

We further analyze the temperature-dependent phase transition in TSI2 by investigating the point-group symmetry of the three crystal structures using the CELLSUB module~\cite{bb2} of Bilbao crystallographic server~\cite{bb1}. There can be three different pathways to $I422$ $\rightarrow$ $C2$ transition via an intermediate phase, which can be $F222$, $I222$, or $I4$ as shown in Fig.~\ref{fig:path}. However, as in our experiments we obtained the $I4$ structure in the intermediate temperature range, this eliminates the possibility first two intermediate phases. During the transition, the point group symmetry lowers from $D_4$ in the $I422$ structure to $C_2$ in $C_2$ phase via $C_4$ of $I4$ structure. Further analysis using the AMPLIMODES module~\cite{amp1,amp2} of the Bilbao Crystallographic Server reveals that in the case of $I422 \rightarrow I4$ transition, it is $\sim$0.007~\AA{} atomic displacements of 8$c$ Wycoff positions of Se atoms which causes the lowering of symmetry from $D_4$ to $C_4$.
This strongly suggests a lattice-driven origin for the observed hidden-order transition near $T^*$ and reinforces the idea that subtle structural instabilities, coupled to electronic degrees of freedom, underpin the emergence of a gapped, correlated ground state well below T$_{CDW}$.

\section{Discussion}

The study of charge density wave (CDW) dynamics and collective excitations in (TaSe$_4$)$_2$I has garnered considerable attention due to its complex band topology\cite{Shi2021,Zahid_Nat.Phy,li2021typeIIIWeyl} and recent claims of axion-like excitations~\cite{AxionGooth2019}. Notably, the realization of a massive phason mode via a Higgs-like mechanism has been proposed to occur exclusively below the secondary transition temperature $T^*$~\cite{bae2025trSTM,kim2023observation}. A recent synchrotron study on TSI2 argued the possibility of Ta-tetramerization and monoclinic distortion only below 120 K\cite{van2001structure}. This was later anticipated in another recent X-ray experiment \cite{Structure_PhysRevLett.87.015502}, further corroborating a symmetry-lowered ground state distinct from the primary CDW phase.

In this context, low-frequency resistance noise provides a powerful window for tracing the underlying fluctuations and emergent order. In broken-symmetry systems exhibiting long-range order, low-frequency fluctuations are often governed by the relaxation dynamics of carriers and defects with finite lifetimes~\cite{kogan1984low,RevModPhysWeissman,CuerrentOpinion_Raychaudhuri}. In non-magnetic conductors, $1/f$ noise typically arises from slow fluctuations in either the carrier density or mobility~\cite{gruner2018density}. Moreover, since the resistance $R$ of a non-magnetic conductor depends on both carrier mobility $\mu$ and carrier density $n$, the normalized variance in resistance fluctuations can be expressed as~\cite{Hooge1994_1byf_noise,hooge1981experimental,RevModPhysWeissman,Kalimuddin.PRL}:

\begin{equation}
\frac{\langle \delta R^2 \rangle}{\langle R^2 \rangle} = 
\frac{\langle \delta n^2 \rangle}{\langle n^2 \rangle} +
\frac{\langle \delta \mu^2 \rangle}{\langle \mu^2 \rangle} +
2 \frac{\langle \delta n \rangle \langle \delta \mu \rangle}{\langle n \rangle \langle \mu \rangle}
\approx \frac{\langle \delta n^2 \rangle}{\langle n^2 \rangle},
\end{equation}

\noindent
where the final approximation holds under the assumption that mobility fluctuations are fast and uncorrelated, averaging out over nanosecond-to-microsecond timescales. In contrast, the correlated and long-timescale resistance fluctuations observed near $T^*$ suggest that the dominant source of noise arises from number density fluctuations.

In our recent study~\cite{Kalimuddin.PRL}, we argued that fluctuations in the condensed carrier density ($n_c$) are transferred to normal carriers ($n_n$) using a two-fluid analogy. At high temperatures, resistance fluctuations are primarily governed by amplitude modes near $T_{\mathrm{CDW}}$, whereas at lower temperatures near $T^*$, the low density of thermally activated quasiparticles points to an increasing contribution from phase fluctuations. The emergence of strong correlations in noise statistics near $T^*$—evident through non-Gaussian probability distributions and enhanced second spectra—further indicates the possible presence of a secondary, symmetry-breaking transition distinct from the conventional CDW order.

It is also noteworthy that our temperature-dependent single-crystal Laue diffraction reveals subtle but distinct anomalies in the lattice parameters (a and c), unit-cell volume, Se–Se dimer bond length, and the dihedral angle between consecutive ${Se_4}$ planes near the $T^*$ (see Fig. S1 of the SM\cite{supply}). These variations indicate a symmetry-lowering distortion concurrent with the onset of enhanced low-frequency noise and thermopower anomalies, suggesting a lattice-driven hidden-order phase. This observation is consistent with previous low-temperature X-ray diffraction studies in (TaSe$_4$)$_2$I\cite{van2001structure,Structure_PhysRevLett.87.015502}.

Moreover, the emergence of the hidden-order phase below $T^* \sim 100~\mathrm{K}$ has profound implications for the axionic electrodynamics proposed in (TaSe$_4$)$_2$I. The observed structural distortion from the $I422$ to $C2$ phase via $I4$ symmetry leads to a renormalized band structure with a small indirect gap ($\sim$0.2~eV), thereby modifying the coupling between the CDW phason and Weyl fermions. This secondary symmetry breaking stabilizes a massive phason mode—a key component of a massive axion-like excitation. Hence, the hidden-order transition to the correlated ground state enhances the robustness and anisotropic character of the axionic CDW state, illustrating the intricate interplay between topology, lattice distortion, and correlated collective excitations in this Weyl-CDW system.

\section{Conclusion}
In summary, we uncover a hidden-order phase transition in the type-III Weyl-CDW semimetal TSI2 at $T^*$ $\sim$100~K—well below the known CDW transition temperature of $T_{CDW}$ $\sim$263~K, using a multifaceted approach that includes electrical resistivity, low-frequency resistance noise spectroscopy, thermoelectric power measurements, and first-principles calculations. This transition is manifested by a pronounced enhancement in the noise exponent ($\alpha$), relative noise variance $\left( \frac{\langle\Delta{R^{2}} \rangle}{\langle R^{2}\rangle}\right)$, and excess kurtosis ($\kappa$). Our Seebeck measurement also displays a sharp anomaly at $T^*$, suggesting a reconstruction of the electronic structure and possible Fermi surface instability. The first-principles calculations reveal that a structural distortion from the high-symmetry $I422$ phase to a low-symmetry $C2$ phase opens a small bandgap of approximately 0.1 eV and lowers the total energy, supporting the emergence of a new gapped, symmetry-broken ground state. Group-theoretical analysis corroborates this transition pathway, highlighting an intermediate $I4$ phase and identifying symmetry-lowering atomic displacements as the driving mechanism. Finally, our findings establish TSI2 as a rare material platform where correlated topological phases and low-dimensional electronic instabilities coexist. The observed hidden-order transition, residing deep within the gapped CDW regime, exemplifies the complex interplay of topology, electron-phonon coupling, and collective excitations. These results not only expand the complex phase diagram of (TaSe$_4$)$_2$I but also open new avenues for exploring emergent secondary order and fluctuation-driven phenomena in topological Weyl-CDW materials.

\section{Acknowledgements}
This work was supported by CSIR-Human Resource Development Group (HRDG) Grant No. (03/ 1511/23/EMR-II) and Department of Science and Technology, Government of India (CRG/2023/001100). The authors thank Riju Pal and Prof. Atindra Nath Pal from SNBNCBS for the technical help. S. Chatterjee thanks Dr. Nitish Mathur for the helpful discussions. Sk Kalimuddin thanks IACS for PhD fellowship. TD and SD acknowledge UGC, India, for support from fellowship. SKP thanks NFSG grant from BITS-Pilani, Dubai campus which supported this research.

\bibliography{Reference}

\newpage
\vspace{2cm}

\begin{center}

\onecolumngrid

\newpage
\textbf{\underline{\large{Supplemental Material}}}\\
\textbf{\large{Emergence of a hidden-order phase well below the charge density wave transition in a topological Weyl semimetal (TaSe$_4$)$_2$I}}\\
\vspace{0.5cm}
Sk Kalimuddin,$^1$ Sudipta Chatterjee,$^2$ Arnab Bera,$^1$ Satyabrata Bera,$^1$ Deep Singha Roy,$^1$ Soham Das,$^1$ Tuhin Debnath,$^1$ Ashis K. Nandy,$^3$ Shishir K. Pandey,$^{4,5,*}$ and Mintu Mondal$^{1,\dagger}$

$^1$\textit{School of Physical Sciences, Indian Association for the Cultivation of Science, Jadavpur, Kolkata 700032, India}\\
$^2$\textit{Department of Condensed Matter and Materials Physics, S. N. Bose National Centre for Basic Sciences, JD Block, Sector III, Salt Lake, Kolkata 700106, India}\\
$^3$\textit{School of Physical Sciences, National Institute of Science Education and Research Bhubaneswar, An OCC of Homi Bhabha National Institute, Khurda Road, Jatni, Odisha 752050, India}\\
$^4$\textit{Department of General Sciences, Birla Institute of Technology and Science, Pilani-Dubai Campus, Dubai International Academic City, Dubai 345055, UAE}\\
$^5$\textit{Department of Physics, Birla Institute of Technology and Science, Pilani, Hyderabad Campus, Jawahar Nagar, Kapra Mandal, Medchal District, Telangana 500078, India}\\
$^*$Author to whom correspondence should be addressed: shishir.kr.pandey@gmail.com; sspmm4@iacs.res.in
\end{center}

\renewcommand{\thesection}{S\arabic{section}}
\renewcommand{\thefigure}{S\arabic{figure}}
\renewcommand{\thetable}{S\arabic{table}}
\setcounter{section}{0}
\setcounter{figure}{0}
\setcounter{table}{0}

\vspace{1.0cm}
\begin{center}
\textbf{Abstract}
\end{center}
This Supplemental Material provides comprehensive experimental details supporting the emergence of a hidden-order phase in the topological Weyl semimetal (TaSe$_4$)$_2$I. We present temperature-dependent single-crystal XRD analysis revealing structural evolution across and well below the charge density wave (CDW) transition, showing distinctive changes at low temperatures. We summarize the robustness of the emergent hidden order phase in different samples and their manifestations in low-frequency noise. The supplementary data further includes extended temperature-dependence of noise exponent ($\alpha$), and excess kurtosis ($\kappa$). It presents a detailed description of the thermo-power measurements using calibrated steady-state methods, including a systematic investigation of thermo-emf response to temperature gradients for multiple samples with different thermal cycles.

\vspace{1cm}

\makeatletter
\let\addcontentsline\origaddcontentsline
\makeatother

\renewcommand{\thesection}{S\arabic{section}}
\renewcommand{\thefigure}{S\arabic{figure}}
\setcounter{section}{0}
\setcounter{figure}{0}

\tableofcontents
\clearpage

{\section{Structural analysis}}
Temperature-dependent single-crystal XRD studies were carried out to understand structural changes with temperature. At room temperature (TaSe$_4$)$_2$I crystallizes in a tetragonal crystal structure (space group \textit{I422}, No. 97). The lattice parameters of the room temperature $\textit{I422}$ structure are found to be a = b = 9.53(7) \AA, c = 12.77(10) \AA; $\alpha=\beta=\gamma=90^{\circ}$. The complete details of the lattice constants, structural parameters, and R factors are listed in Table \ref{Table:Structure}. The RT crystal structure is shown in Figure 1 of the main manuscript.

\begin{table}[htbp]
\centering
\caption{Single crystal XRD analysis}
\label{tab:xrd}
\begin{tabular}{||c|cc|cc|cc|cc|cc||}
\hline\hline
\multirow{2}{*}{Temp (K)} 
& \multicolumn{2}{c|}{\textit{I422}} 
& \multicolumn{2}{c|}{\textit{I4}} 
& \multicolumn{2}{c|}{\textit{I222}} 
& \multicolumn{2}{c|}{\textit{F222}} 
& \multicolumn{2}{c||}{\textit{C2}} \\
\cline{2-11}
& $R_1$ (\%) & $wR_1$ (\%) 
& $R_1$ (\%) & $wR_1$ (\%) 
& $R_1$ (\%) & $wR_1$ (\%) 
& $R_1$ (\%) & $wR_1$ (\%) 
& $R_1$ (\%) & $wR_1$ (\%) \\
\hline
 &   &   &   &   &   &   &   &   &   &   \\
300 & 2.06 & 5.18 & NA & NA & NA & NA & NA & NA & NA & NA \\
 &   &   &   &   &   &   &   &   &   &   \\
\hline
 &   &   &   &   &   &   &   &   &   &   \\
270 & 7.51 & 20.22 & NA & NA & NA & NA & NA & NA & NA & NA \\
 &   &   &   &   &   &   &   &   &   &   \\
\hline
 &   &   &   &   &   &   &   &   &   &   \\
220 & 7.80 & 20.20 & 7.55 & 19.62 & NA & NA & NA & NA & NA & NA \\
 &   &   &   &   &   &   &   &   &   &   \\
\hline
 &   &   &   &   &   &   &   &   &   &   \\
175 & 6.02 & 16.84 & 4.43 & 12.58 & NA & NA & NA & NA & NA & NA \\
 &   &   &   &   &   &   &   &   &   &   \\
\hline
 &   &   &   &   &   &   &   &   &   &   \\
95  & NA & NA & 4.41 & 12.50 & NA & NA & NA & NA & NA & NA \\
 &   &   &   &   &   &   &   &   &   &   \\
\hline
\end{tabular}
\label{Table:Structure}
\end{table}

\begin{figure}[h]
\centering
\includegraphics[width=1.0\columnwidth]{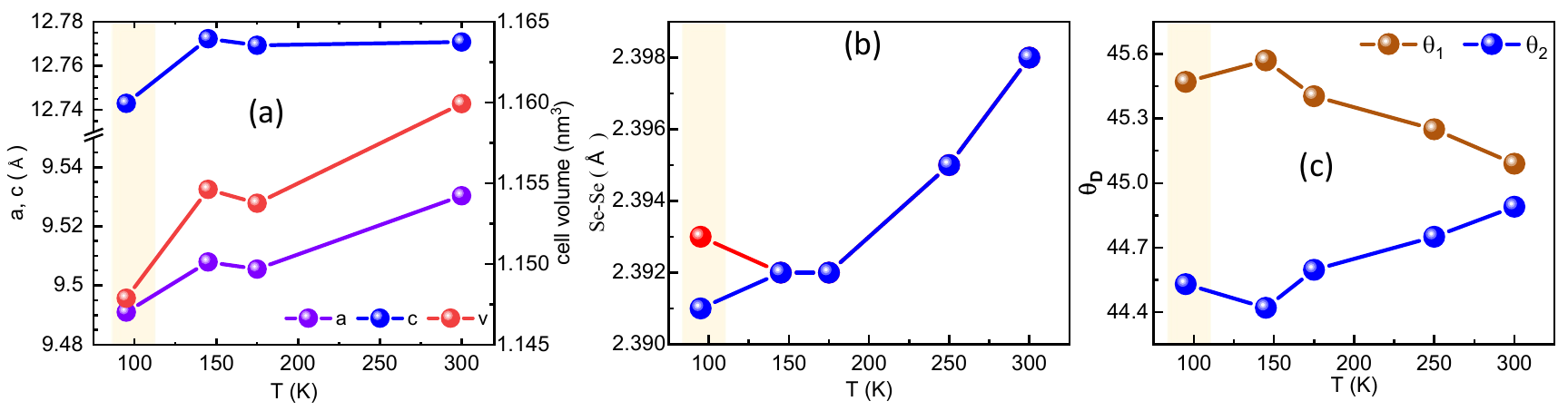}
\caption{Temperature dependence of various physical parameters: (a) (left) lattice parameter $a,c$ (in \AA), (right) unit cell volume $V$ (in \AA$^3$), (b) Selenium-Selenium dimer bond-length (in \AA), (c) di-hedral angle between two consecutive $Se_4$ planes over the temperature range 90–300~K.}
\label{Fig:Structure}
\end{figure}

\newpage
{\section{Noise measurements in sample-II}}

\begin{figure}[h!]
\centering
\includegraphics[width=1.0\columnwidth]{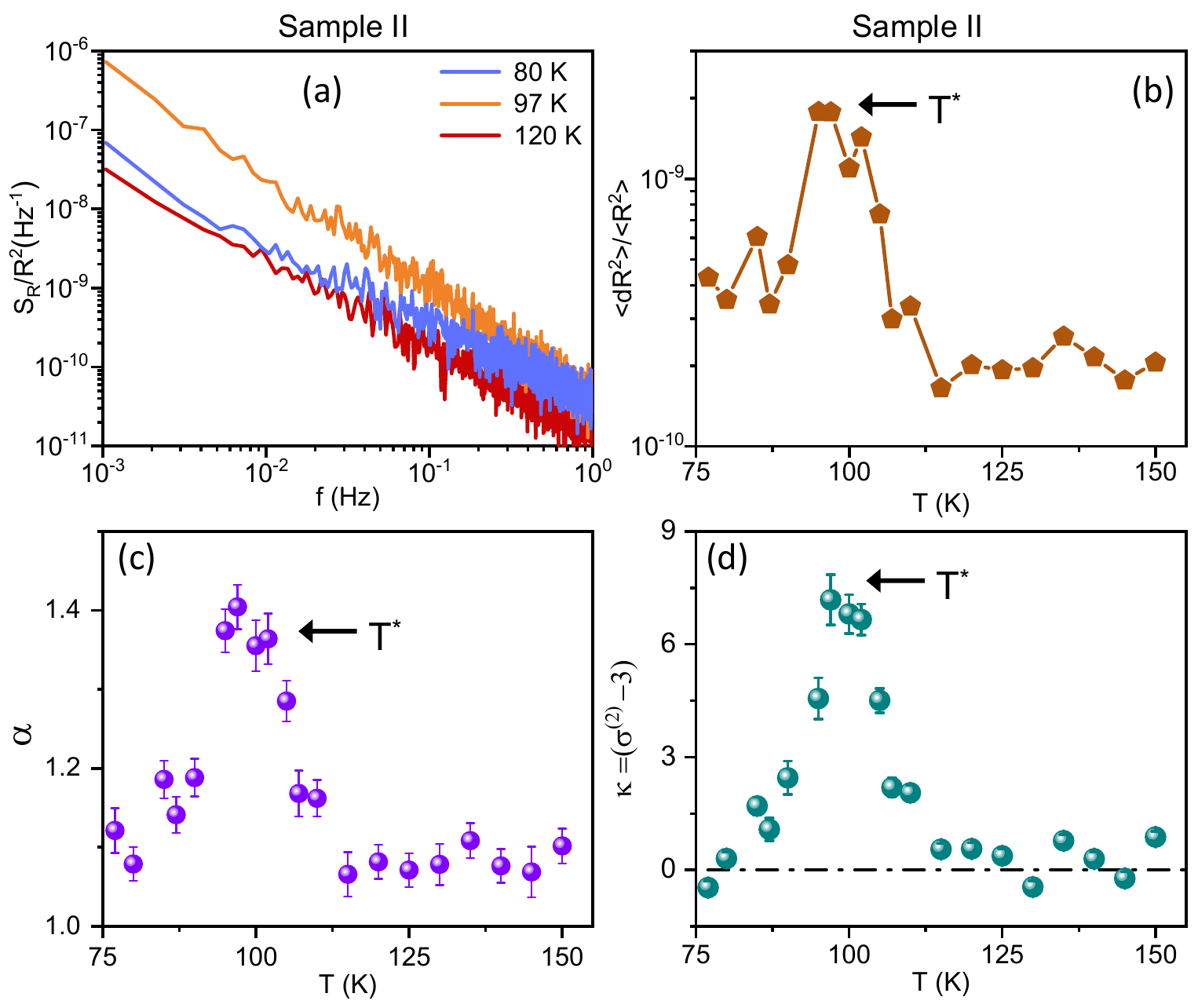}
\caption{{\em{Noise characteristics of Sample II.}}(a) Frequency-dependent resistance noise power spectral density \(S_R/R^2\) (in Hz\(^{-1}\)) measured at selected temperatures (80~K, 97~K, 120~K). (b) Temperature dependence of the relative noise power, showing a distinct peak near a characteristic temperature \(T^*\). (c) The exponent \(\alpha\) of the frequency dependence \(S_R/R^2 \sim 1/f^\alpha\) as a function of temperature. (d) Temperature dependence of the normalized fourth moment \(\langle \sigma^{(2)} \rangle - 3\), indicating non-Gaussian noise statistics across the measured temperature range.} 
\label{Fig:NoiseII}
\end{figure}

\clearpage
{\section{Thermoelectric power}}
The Seebeck coefficient (S), a fundamental parameter in thermoelectric materials, quantifies the voltage generated in response to a temperature gradient across a material. Accurate measurement of this coefficient as a function of temperature is crucial for evaluating and optimizing materials for thermoelectric applications. Here, we employed the steady-state method similar to our previous work\cite{Sruthi2022}, where we ensured a steady temperature gradient across the sample and measured the resulting thermoelectric voltage using a nano-voltmeter(Keithley 2182A). The Seebeck coefficient is determined from the ratio of the voltage difference to the temperature difference\cite{Patel04072017}. The experimental setup for the Seebeck coefficient has been shown in Figure \ref{fig:Calibration}(a).

\vspace{2.0cm}
\subsection{Calibration of Seebeck setup}
\begin{figure}[h]
\centering
\includegraphics[width=1.0\columnwidth]{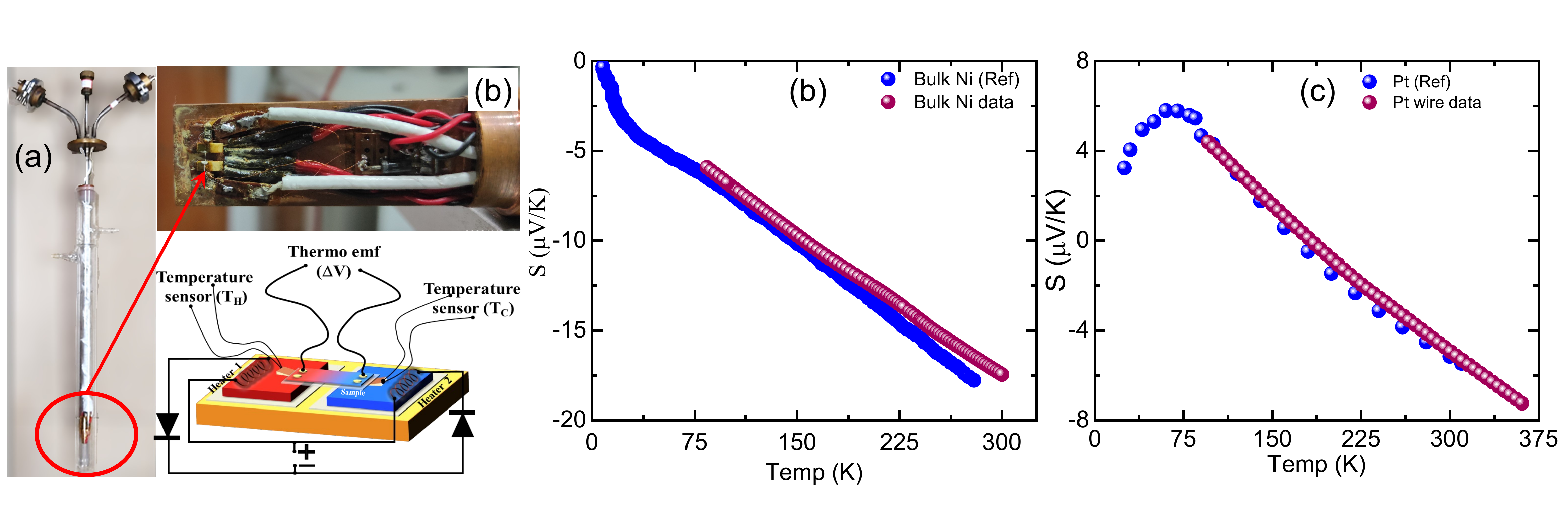}   	
\caption{\textbf{Thermoelectric power setup and calibration:} (a) Liquid nitrogen-based cryogenic insert for thermo-emf measurement. (b) (top) Image of the sample holder with (TaSe$_4$)$_2$I mounted on it, and (bottom) Wiring configuration for local heating and measurement of thermo-electric voltage. (b) Experimental Seebeck(purple) vs reference data\cite{Seebeck_Nickel} (blue) for Nickel. (c) Experimental Seebeck(purple) vs reference data\cite{Seebeck_Platinum} (blue) for platinum wire.}
\label{fig:Calibration}	
\end{figure}

The absolute calibration of the Seebeck has been done using a high-quality (99.99 \% pure ) Nickel piece and a platinum wire of purity 99.99 \%. Our experimentally measured Seebeck coefficient over a temperature window of 80-375~K matches well with reference data for both Nickel\cite{Seebeck_Nickel} and Platinum\cite{Seebeck_Platinum} respectively. The experimental Seebeck data vs reference data plot for Ni and Pt wire are shown in Figure \ref{fig:Calibration}(b-c) respectively.

\section{Thermoelectric power of (TaSe$_4$)$_2$I}
The thermoelectric power measurement of bulk (TaSe$_4$)$_2$I single crystal has been done in both heating and cooling cycles. The S vs T for heating and cooling cycles are plotted in Figure \ref{Fig:Seebeck}(a-b) respectively. The trend in the S vs T plot is linear for temperatures above $T_{CDW}$, and an approximate $\frac{1}{T}$ dependence below  $T_{CDW}$.

\subsection{Thermo-emf vs temperature gradient}
\begin{figure}[h!]
\centering
\includegraphics[width=1.0\columnwidth]{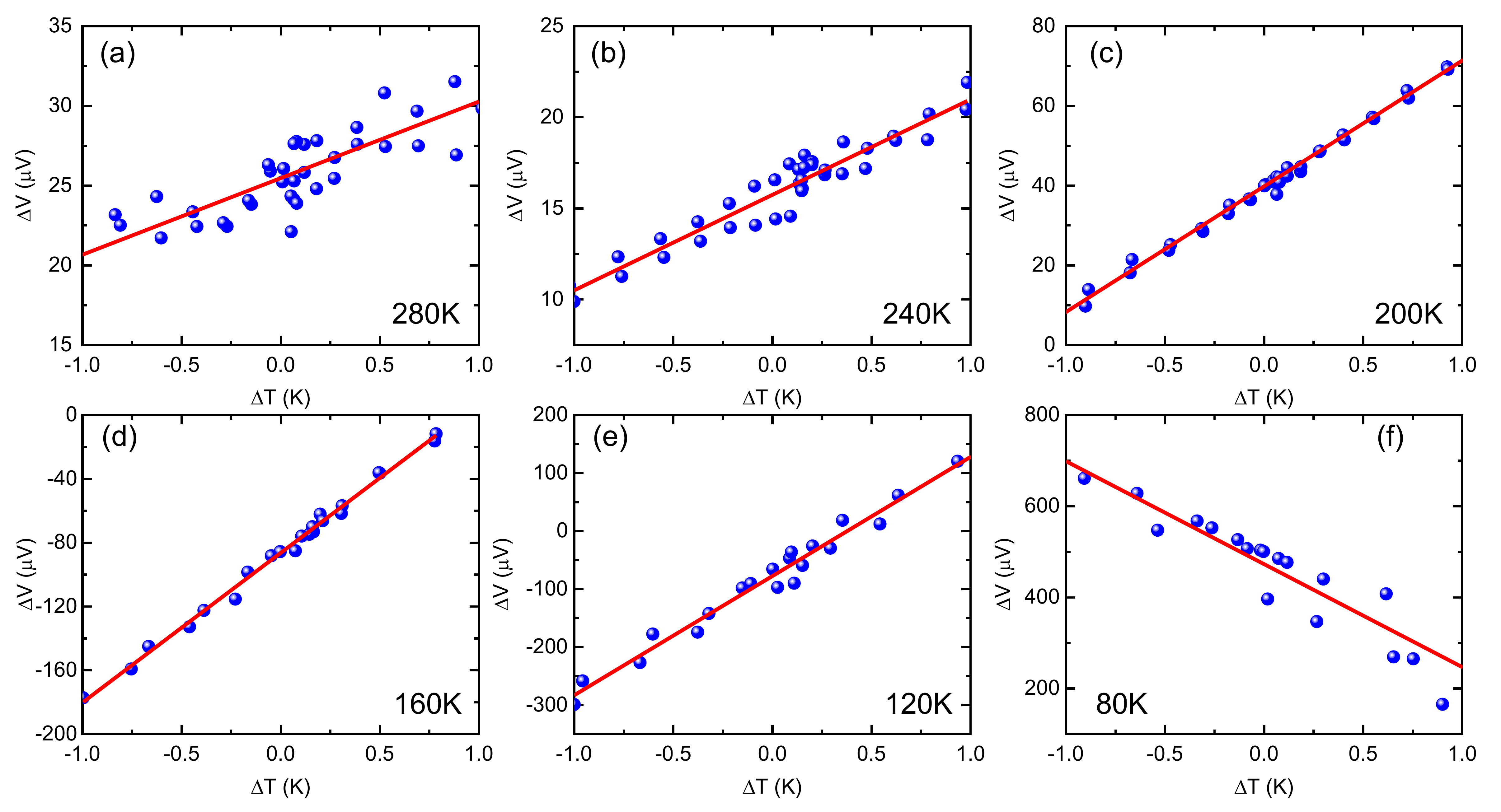}   	
\caption{\textbf{Thermo-emf vs temperature gradient:} (a-f) Plots of thermo-electric voltages corresponding to temperature gradients at six representative temperatures.}
\label{fig:Slopes}	
\end{figure}

Seebeck voltage varies linearly with the magnitude of the temperature gradient across the sample. Plots of thermo-electric voltages corresponding to temperature gradients at six representative temperatures have been shown in Figure \ref{fig:Slopes}(a-f). 

\newpage

\subsection{Temperature dependence of Seebeck coefficient}
\begin{figure}[h!]
\centering
\includegraphics[width=1.0\columnwidth]{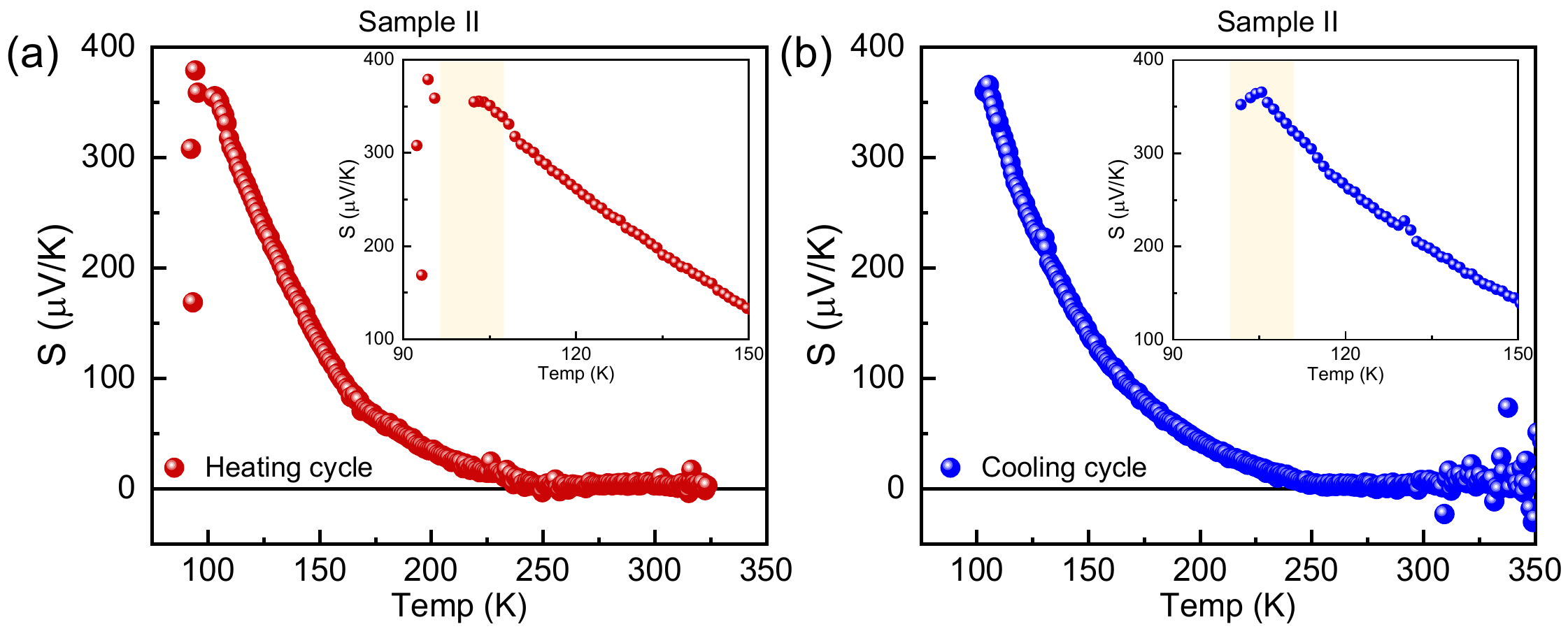} 
\caption{\textbf{Extended data for thermo-emf:} Seebeck coefficient(S) vs temperature(K) of (TaSe$_4$)$_2$I sample II in (a) heating and (b) cooling cycles, respectively. Inset is a zoomed-in plot to highlight the anomaly near $T^*\sim100~K$}
\label{Fig:Seebeck}	
\end{figure}

\end{document}